\documentclass[12pt]{article}
\usepackage{latexsym}
\newtheorem{theorem}{Theorem}

\newcommand{\proof}{\noindent{\bf Proof.}\ }
\begin{document}

\newcommand{\lap}{\bigtriangleup}
\def\be{\begin{equation}}
\def\ee{\end{equation}}
\def\bea{\begin{eqnarray}}
\def\eea{\end{eqnarray}}
\def\beas{\begin{eqnarray*}}
\def\eeas{\end{eqnarray*}}
\def\n#1{\vert #1 \vert}
\def\nn#1{{\Vert #1 \Vert}}
 
\def\R{{\rm I\kern-.1567em R}}
\def\N{{\rm I\kern-.1567em N}}
 
\def\supp{\mbox{\rm supp}\,}
\def\vol{\mbox{\rm vol}\,}
\def\dt{\partial_t}

\def\ekin{E_{\rm kin}}
\def\epot{E_{\rm pot}}

\def\C{{\cal C}}
\def\F{{\cal F}}
\def\G{{\cal G}}
\def\H{{\cal H}}
\def\Hc{{\cal H_C}}
\def\Hr{{{\cal H}_r}}
\def\M{{\cal M}}
\def\r{\rho}
\def\e{\epsilon}

\def\prfe{\hspace*{\fill} $\Box$

\smallskip \noindent}

\title{Non-linear stability of gaseous stars}

\author{ Gerhard Rein\\
Institut f\"ur Mathematik der
Universit\"at Wien\\
Strudlhofgasse 4\\
1090 Vienna, Austria}
\date{}
\maketitle

\begin{abstract}
We construct steady states of the Euler-Poisson system
with a barotropic equation of state as minimizers
of a suitably defined energy functional.
Their minimizing property implies the non-linear stability
of such states against general, i.e., not necessarily spherically symmetric
perturbations. The mathematical approach is based on
previous stability results for the Vlasov-Poisson system
by Y.~Guo and the author, exploiting the energy-Casimir technique.
The analysis is conditional in the sense that it {\em assumes} 
the existence of solutions to the initial value problem for the Euler-Poisson
system which preserve mass and energy.
The relation between the Euler-Poisson and the Vlasov-Poisson
system in this context is also explored.
  
\end{abstract}

\section{Introduction}
\setcounter{equation}{0}

Consider a self-gravitating fluid ball in $\R^3$
where the fluid has mass density $\rho = \rho(t,x)\geq 0$ and velocity field
$u=u(t,x)\in \R^3$; $t\geq 0$ denotes time and $x\in\R^3$ position.
In the simplest case such a system obeys the Euler-Poisson equations
\be \label{rhocont}
\dt \rho + \nabla\cdot(\rho u) = 0,
\ee
\be \label{euler}
\rho \dt u + \rho (u\cdot \nabla) u = - \nabla p - \rho \nabla V,
\ee
\be \label{poisson}
\lap V = 4 \pi \rho,\ \lim_{|x|\to \infty} V(t,x) = 0,
\ee
where $V=V(t,x)$ denotes the self-consistent gravitational potential
and $p=p(t,x)$ is the pressure in the fluid. To close this system
an equation of state needs to be supplied
which relates the pressure and the density:
\be \label{eqstate}
p=P(\rho).
\ee
An equation of state where the pressure does not 
depend on the temperature
or specific entropy is often referred to as barotropic. More sophisticated
models for a star would contain equations for the temperature
and for the thermonuclear processes within the star,
but in astrophysics the present model is often used as a
simple description of a gaseous star,
in particular if---as in the 
present note---the issue is the stability of steady states
since this problem is non-trivial already for this simple model.

To approach this problem we define the energy
of a state $(\rho,u)$ as
\be \label{energy}
\H(\rho,u):= \frac{1}{2} \int |u|^2 \rho dx + \int \Phi(\rho)\, dx
-\frac{1}{2}\int \frac{\rho(x) \rho(y)}{|x-y|}dy\,dx.
\ee
Here $\Phi$ is defined in such a way that for $\rho>0$, 
\be \label{PhiP}
P'(\rho) = \rho\,\Phi''(\rho).
\ee
Under appropriate assumptions on $\Phi$ or $P$ respectively,
which include---but are much more general than---equations 
of state of the form $P(\rho)= c \rho^\gamma$
with $\gamma >4/3$,
this functional has a minimizer under the mass constraint
\[
\int \rho \, dx = M
\]
with $M>0$ prescribed. This follows from a result of the author
established in \cite{R3}. Such a minimizer
is a spherically symmetric steady state of the Euler-Poisson system
with vanishing velocity field and as such is
referred to as a static solution of the system, as opposed to
the more general concept of a stationary solution, 
which is also time-independent but may have non-vanishing velocity field.
From the fact that the steady state minimizes the energy
one can deduce a rigorous, non-linear stability result.
Since in order to solve
the variational problem stated above we need to make no symmetry 
assumption  we obtain stability against
general, i.e., not necessarily spherically symmetric perturbations,
even though the steady state obtained as an energy minimizer must 
a posteriori be spherically symmetric.
An essential point for the stability analysis is that both the
energy defined in (\ref{energy}) and the total mass $\int \rho\, dx$
are conserved along solutions of the time dependent problem.
Since the initial value problem
for the Euler-Poisson system is far from being completely understood
and one has to expect the occurrence of shocks or gravitational
collapse our stability result is conditional in the following sense:
As long as a solution to the initial value problem exists and preserves
mass and energy it satisfies the stability estimate.

The paper proceeds as follows: In the next section we study
the variational problem stated above. 
Then we show that a minimizer 
is a spherically symmetric, static solution. In Section~3 
we carry out the stability analysis.
In a final section we discuss the relation of the present result
to those obtained by Y.~Guo and the author for the Vlasov-Poisson
system
\[
\partial_t f + v \cdot \nabla_x f - \nabla V \cdot 
\nabla_v f = 0 ,
\]
\[ 
\lap V = 4 \pi\, \r,\ \lim_{|x| \to \infty} V(t,x) = 0 , 
\]
\[
\r(t,x)= \int f(t,x,v)dv.
\]
Here the dynamic variable is the number density in phase space,
$f=f(t,x,v)$, of an ensemble of massive particles  
with spatial density $\rho$, interacting by the gravitational
potential $V$ which the particles create collectively;
$v\in\R^3$ denotes the momentum or velocity coordinate in phase space.
In this model collisions among the particles are neglected
which is appropriate if one is describing a galaxy or 
a globular cluster.  
In \cite{G1,G2,GR1,GR2,GR3,R1,R2} a variational approach to
the question of the stability of steady states of this system
was developed, which relied on the minimization of appropriately
defined energy-Casimir functionals. In \cite{R3}
the author observed that these functionals, which act on
phase space densities $f=f(x,v)$, can in a natural way be
reduced to functionals acting on spatial densities $\rho=\rho(x)$, 
and he studied the variational problem for these reduced
functionals, without realizing that
this yields the stability result for the Euler-Poisson
problem which we discuss here.
Incidentally, the close relation between the stability
of barotropic stars and of stellar systems was already observed on a
formal level in the astrophysics literature and is often referred
to as Antonov's First Law, cf.\ \cite[5.2 (b)]{BT}. 

We conclude this introduction with some additional references
to the literature. A lot of background on the physics of stars
can be found in \cite{Chan} and \cite{Kipp}.
An excellent and broad overview of mathematical results 
for hydrodynamical models
of gaseous stars is given in \cite{Duc} which also contains 
many further references. 
Our variational approach is related to  
the concentration-compactness principle
due to P.-L.~Lions \cite{L}.
The use of
energy-Casimir functionals for questions of stability was discussed in
a very broad context in \cite{HMRW}.
As aimed more specifically to the existence and stability
properties of steady states for gaseous stars we mention
\cite{DLYY,Lin} and the references there.

\section{Energy minimizers and steady states}
\setcounter{equation}{0}

If we minimize the energy $\H$ as defined in (\ref{energy})
over all ``relevant'' states $(\rho,u)$ then a minimizer will clearly
satisfy $u=0$. Thus we study the following variational problem:
For a density $\rho$ we define the induced potential
\[
V_\rho(x) := - \int \frac{\rho(y)}{|x-y|} \, dy,
\] 
and the potential energy
\be \label{potendef}
\epot(\rho) :=
- \frac{1}{2} \int\!\!\!\int \frac{\rho(x) \rho(y)}{|x-y|}dy\,dx
= \frac{1}{2} \int \rho\,V_\rho  dx
= - \frac{1}{8 \pi} \int|\nabla V_\rho|^2 dx.
\ee
We want to minimize the functional
\be \label{redenergy}
\Hr(\rho)
:= \int \Phi(\r(x))\, dx + \epot(\r)
\ee
over the constraint set
\be \label{spacedef}
\F_M := \Bigl\{ \r \in L^1 (\R^3) 
\mid
\rho\geq 0,\ \int \Phi(\r) < \infty, \ \int \r = M \Bigr\}
\ee
for $M>0$ prescribed
and $\Phi$ satisfying the following

\smallskip
\noindent {\bf Assumptions on $\Phi$}: 
$\Phi \in C^1 ([0,\infty[)$, $\Phi(0)=0=\Phi'(0)$, and
\begin{itemize}
\item[$(\Phi 1)$]
$\Phi$ is strictly convex. 
\item[$(\Phi 2)$]
$\Phi (\r) \geq C \r^{1+1/n}$ for $\r \geq 0$ large,  with $0 < n < 3$,
\item[$(\Phi 3)$]
$\Phi (\r) \leq C \r^{1+1/n'}$ for $\r \geq 0$ small,  with $0 < n' < 3$.
\end{itemize} 

Note that $\rho \in L^1\cap L^{4/3}(\R^3)$
and hence $V_\rho \in L^p(\R^3)$, $\nabla V_\rho \in L^q(\R^3)$
for $\rho \in \F_M$ and
any $p\in ]3,12]$ and $q\in ]3/2,12/5]$ so that the potential energy
is well defined.
The following result was proved earlier by the author,
cf.\ \cite[Thm.~3.1]{R3}; it is included here for easier reference.
\begin{theorem} \label{ccp}
The functional $\Hr$ is bounded from below on $\F_M$
with $\inf_{\F_M}\Hr < 0$.
Let $(\r_i)\subset \F_M$ be a minimizing sequence 
of $\Hr$. Then there exists a sequence of shift vectors 
$(a_i)\subset \R^3$ and a subsequence, again denoted
by $(\r_i)$, such that for any $\epsilon >0$ there exists $R>0$
with
\[
\int_{a_i + B_R} \r_i (x)\, dx \geq M - \epsilon,\ i \in \N,
\]
\[
T \r_i := \r_i(\cdot + a_i) \rightharpoonup \r_0
\ \mbox{weakly in}\ L^{1+1/n}(\R^3),\ i \to \infty,
\]
and
\[
\int_{B_R} \r_0 \geq M - \epsilon .
\]
Finally,
\[
\nabla V_{T\r_i} \to \nabla V_0\ 
\mbox{strongly in}\ L^2(\R^3),\ i \to \infty,
\] 
where $V_0=V_{\rho_0}$,
and $\r_0 \in \F_M$ is a minimizer of $\Hr$.
\end{theorem}
Here and in the following we denote for $R>0$,
\[
B_R 
:= \{x \in \R^3 | \n{x} \leq R\}.
\]
 
{\bf Remark 1.} As noted in the introduction, the function $\Phi$ appearing
in the definition of the energy and the function $P$ defining the equation
of state (\ref{eqstate}) have to satisfy the relation (\ref{PhiP}).
If we think of $P$ as the basic, given quantity then we define
\[
\Phi(\rho):= \int_0^\rho\int_0^\sigma 
\frac{P'(\tau)}{\tau}d\tau\,d\sigma,\ \rho \geq 0,
\]
and this function satisfies the assumptions $(\Phi 1), (\Phi 2),\ (\Phi 3)$
provided $P'(\tau)/\tau$ is integrable on $]0,\infty[$
and the assumptions
\begin{itemize}
\item[(P1)]
$P'>0$, 
\item[(P2)]
$P' (\tau) \geq C \tau^{1/n},\ \tau \geq 0$ large,  with $0 < n < 3$,
\item[(P3)]
$P' (\tau) \leq C \tau^{1/n'},\ \tau \geq 0$ small,  with $0 < n' < 3$
\end{itemize}   
hold. This is the case in particular for polytropic equations of state
\[
P(\rho)= c \rho^\gamma
\]
with $c>0$ and $\gamma > 4/3$, in which case 
\[
\Phi(\rho) = \frac{c}{\gamma - 1} \rho^\gamma.
\]
Equations of state where the dependence of the pressure on the density
is different for large and for small values of the density are motivated
from a physics point of view.
\smallskip

{\bf Remark 2.} The need for the spatial shifts in Theorem~\ref{ccp}
is not technical since without them the assertion of
the theorem is false: Given a minimizer $\rho_0$ the
fact that the functional $\Hr$ is translation invariant implies
that by shifting $\rho_0$
off to infinity one obtains a minimizing sequence which
converges weakly to zero and not to a minimizer.
\smallskip 

Now that a minimizer is obtained we show that it is a steady state
of the Euler-Poisson system;
in order to avoid technical discussions concerning regularity
let us now assume in addition to the above that $\Phi \in C^2(]0,\infty[)$
which means $P \in C^1(]0,\infty[)$.   
\begin{theorem} \label{minim=ss}
Suppose that $\rho_0 \in \F_M$ is a minimizer of the functional
$\Hr$ with induced potential $V_0$.
Then there exists a Lagrange multiplier $E_0\leq 0$ such that
\[
\r_0 = \left\{ 
\begin{array}{ccl}
(\Phi')^{-1}(E_0 - V_0)&,& V_0 < E_0, \\
0 &,& V_0 \geq E_0. 
\end{array}
\right.
\]
The functions $\rho_0$ and $V_0$ are spherically symmetric with respect to
some point in $\R^3$,
as functions of the radial variable $\rho_0$ is decreasing and 
$V_0$ increasing, $\rho_0 \in C(\R^3)$ with finite mass $M$,
$V_0 \in C^2(\R^3)$ with $\lim_{|x|\to \infty} V_0(x)=0$,
and $(\rho_0,u_0=0)$ is a time-independent solution
of the Euler-Poisson system with equation of state
\[
P(\rho)= \int_0^\rho \sigma \Phi''(\sigma)\, d\sigma.
\]
If $\Phi'(\rho)\leq C\rho^{1/n'}$ for $\rho\geq 0$ small
then $E_0<0$ and $\rho_0$ has compact support.
\end{theorem}
The assumptions on $\Phi$ imply that $\Phi'$ is invertible
on $[0,\infty[$. The growth condition on $\Phi'$ implies
the condition $(\Phi 3)$ but is not equivalent to it.

\noindent
{\bf Proof of Theorem~\ref{minim=ss}.}
We start by deriving the Euler-Lagrange equation for the variational problem.
Let $\rho_0\in \F_M$ be a minimizer with induced potential $V_0$.
For $\e > 0$ define 
\[
S_\e :=\{x\in\R^3 | \e \leq \rho_0(x) \leq 1/\e \};
\]
think of $\rho_0$ as a pointwise defined representative of the minimizer.
For a test function $w\in L^\infty(\R^3)$ which has compact support
and is non-negative on $\R^3\setminus S_\e$ define for $\tau\geq 0$ small,
\[
\rho_\tau := \rho_0 + \tau\, w
- \tau \frac{\int w\, dy}{\vol S_\e} {\bf 1}_{S_\e},
\]
where ${\bf 1}_{S_\e}$ denotes the indicator function of the set $S_\e$.
Then $\rho_\tau \geq 0$ and $\int \rho_\tau = M$ so that $\rho_\tau \in \F_M$
for $\tau \geq 0$ small. Since $\rho_0$ is a minimizer of $\Hr$,
\[
0 \leq
\Hr(\rho_\tau) - \Hr(\rho_0) =
\tau \int (\Phi'(\rho_0) + V_0)
\left(w - \frac{\int w\,dy}{\vol S_\e} {\bf 1}_{S_\e}\right)\, dx
+ {\rm o}(\tau).
\]
Hence the coefficient of $\tau$ in this estimate must be non-negative,
which we can rewrite in the form
\[
\int \left[\Phi'(\rho_0) + V_0 - \frac{1}{\vol S_\e} 
\left(\int_{S_\e}(\Phi'(\rho_0) + V_0)\,dy \right)\right]\, w\, dx \geq 0.
\]
This holds for all test functions $w$ as specified above, and hence
$\Phi'(\rho_0) + V_0 = E_\e$ on $S_\e$ and $\Phi'(\rho_0) + V_0 \geq E_\e$
on $\R^3\setminus S_\e$ for all $\e >0$ small enough. 
Here $E_\e$ is some constant which by the first relation must be independent 
of $\e$, and taking $\e \to 0$ proves the relation between $\rho_0$ and $V_0$.

The symmetry assertion follows by a rearrangement argument: 
Let $\rho_0^\ast$ denote the symmetric decreasing rearrangement
of $\rho_0$. Then 
\[
\int \Phi(\r_0) = \int \Phi(\r_0^\ast),\
\epot (\r_0) \geq \epot(\r_0^\ast),
\] 
and $\r_0^\ast \in \F_M$, cf.\ \cite[Thms.\ 3.7, 3.9]{LL}.
Since $\rho_0$ is a minimizer of $\Hr$ equality must hold
in the estimate for the potential energy which implies that
$\rho_0$ must be spherically symmetric with respect to some point in $\R^3$,
and decreasing as a function of the radial variable. 
Alternatively, one could use \cite[Thm.~4]{GNN}
to conclude the spherical symmetry.

We now continue essentially as in the proof of \cite[Thm.~3]{R2}.
First we observe that by assumptions $(\Phi 1)$ and $(\Phi 2)$
and the mean value theorem
\[
\Phi'(\rho) \geq \Phi'(\sigma) = \frac{\Phi(\rho)-\Phi(0)}{\rho - 0}
\geq C \rho^{1/n}
\]
for all $\rho$ large, with some intermediate value $0\leq \sigma \leq \rho$.
Hence
\[
\rho_0(x) \leq C(1+(E_0-V_0(x))^n)
\]
and since $V_0 \in L^{12}(\R^3)$ and $n<3$ we have $\rho_0 \in L^1\cap L^4(\R^3)$.
For any $R>1$ we split the convolution
integral for $V_0$ according to $\n{x-y} < 1/R$, $1/R \leq \n{x-y} < R$,
and $\n{x-y} \geq R$ to obtain
\[
- V_0 (x) \leq
C \nn{\rho_0}_4 \left( \int_0^{1/R} r^{2-4/3}dr\right)^{3/4}
+
R \int_{\n{y} \geq \n{x} - R} \rho_0(y)\, dy +\frac{M}{R},\ 
\n{x} \geq R.
\]
This implies that
$V_0 \in L^\infty (\R^3)$ with $V_0 (x) \to 0,\ \n{x} \to \infty$.
The asserted regularity of $\rho_0$ and $V_0$ now follows from
the relation between these two quantities and Sobolev's embedding theorem.

The limiting behavior of $V_0$ and the relation between
$\rho_0$ and $V_0$ implies that $E_0 \leq 0$, since otherwise
$\rho_0 (x) \geq (\Phi')^{-1}(E_0/2)>0$ for $|x|$ large which would
contradict the integrability of $\rho_0$.

Next we check that $(\rho_0, u_0=0)$ solves the stationary
Euler-Poisson system with the asserted equation of state.
Since by construction $V_0$ is the potential induced by $\rho_0$
all that needs to be checked is the Euler equation which takes the form
\[
\nabla p_0 + \rho_0 \nabla V_0 = 0.
\]
Now $\Phi'(\rho_0) = E_0 - V_0$ on $\supp \rho_0$. Taking the gradient
of this relation proves the assertion, provided $P' = \Phi'' \rho$.

It remains to show that under the additional growth condition on $\Phi'$
the density is compactly supported and the cut-off energy $E_0$
strictly negative.
Since $V_0$ increases to zero at spatial infinity
the relation between $\rho_0$ and $V_0$ implies that $\rho_0$
has compact support provided $E_0<0$. If we assume that $E_0=0$
then the additional assumption on $\Phi'$ implies that
\[
\rho_0(x) \geq C (-V_0(x))^{n'}
\]
for $|x|$ large, i.e., $\rho_0$ small. But a simple expansion of the
Green's function in the convolution formula for $V_0$
and the fact that $\int \rho_0 = M$ show that
$-V_0(x) \geq M/(3|x|)$ for $|x|$ large. Inserting this into the
estimate for $\rho_0$ from below yields a contradiction to the integrability 
of $\rho_0$. Hence $E_0<0$, and the proof is complete. \prfe

In view of the stability result discussed in the next section
it would be desirable to know that for fixed $M$ there is up to
spatial shifts a
unique minimizer of $\Hr$ in $\F_M$ or at least that the
minimizers are isolated up to spatial shifts. Numerical evidence
obtained by solving the spherically symmetric Poisson problem
$\lap V_0 = 4 \pi \rho_0 = 4 \pi (\Phi')^{-1}(E_0-V_0)$
seems to indicate that the minimizers are in general not unique
but are isolated.
An example where the minimizer is unique is provided by the polytropic
equation of state:
\smallskip

{\bf Remark 3.} Let $P(\rho)=c \rho^{\gamma}$, i.e., 
$\Phi(\rho)=\frac{c}{\gamma-1} \rho^\gamma$ with $\gamma > 4/3$.
Then for every $M>0$ there exists up to spatial shifts exactly
one minimizer $\rho_0 \in \F_M$ of $\Hr$.
This follows from the fact that $E_0 - V_0$
solves the Emden-Fowler equation
\[ 
\frac{1}{r^2} (r^2 z')' = - c z_+^n,\ r>0, 
\]
where $1+1/n=\gamma$ and $z_+$ denotes the positive part of $z$. 
All solutions of this equation which are regular 
at the center are related by a scaling transformation, and
prescribing the mass fixes the corresponding scaling parameter.
For more details cf.\ \cite[Thm.~3 (b)]{R2}.

\section{The stability analysis}
\setcounter{equation}{0}

We start by expanding the total energy $\H$ about a minimizer $\rho_0$:
For $\rho \in \F_M$ and $u$ such that $\int |u|^2 \rho < \infty$ we have
\be  \label{d-d}
\H (\rho,u)- \H (\rho_0,0)
= d(\rho,\rho_0)-\frac{1}{8 \pi}
\|\nabla V_\rho-\nabla V_0\|^2_2 + \frac{1}{2} \int |u|^2 \rho dx
\ee
where
\[
d(\rho,\rho_0) := \int \Bigl[\Phi(\rho)-\Phi(\rho_0) +
(V_0 - E_0)(\rho-\rho_0)\Bigr]\,dx;
\]
we are allowed to insert the term $E_0\,(\rho - \rho_0)$ into $d$ 
since $\int \rho = \int \rho_0$.
The fact that $\Phi$ is convex and the relation between
$\rho_0$ and $V_0$ established in Theorem~\ref{minim=ss} imply that
for all $\rho \in \F_M$, 
\[ 
d(\rho,\rho_0) \geq 0.
\]
Moreover, $d(\rho,\rho_0)=0$ only if $\rho=\rho_0$, and if one assumes in 
addition that $\Phi''$ is bounded away from zero then
\[
d(\rho,\rho_0) \geq C \nn{\rho-\rho_0}_2^2,\ \rho \in \F_M.
\]
What is irritating about (\ref{d-d}) is the minus sign in front
of the difference term for the gravitational fields, but this
term converges to zero along minimizing sequences, cf.\ Theorem~\ref{ccp}.

Our stability analysis relies on the fact that the mass and the
total energy $\H$ defined in (\ref{energy}) are conserved along
solutions of the time-dependent Euler-Poisson system.
Formally, conservation of mass follows from the continuity 
equation (\ref{rhocont}), and conservation of energy follows 
by a straight forward computation from
all three equations (\ref{rhocont}), (\ref{euler}), (\ref{poisson})
together, provided the relation (\ref{PhiP}) between the equation of state 
$p = P(\rho)$ and the function $\Phi$ which appears in the energy
holds. However, from a rigorous mathematical point of view 
insisting on these conservation laws amounts to an {\em assumption}
on the solutions of the initial value problem. This is reflected
in the statement of our stability result. 
\begin{theorem} \label{stability}
Let $\rho_0 \in \F_M$ be a minimizer of $\Hr$ with induced potential
$V_0$, and assume that it is unique up to spatial shifts.
Then for every $\e >0$ there is a $\delta>0$ such that the following holds:
For every solution $t \mapsto (\rho(t),u(t))$ of the Euler-Poisson system
with $\rho(0) \in \F_M$ which preserves mass and energy
the initial estimate
\[
d(\rho(0),\rho_0) + \frac{1}{8\pi} \|\nabla V_{\rho(0)}-\nabla V_0\|_2^2
+ \frac{1}{2} \int |u(0)|^2 \rho(0)\,dx < \delta
\]
implies that for every $t \geq 0$ for which the solution still exists
there is a shift vector $a \in \R^3$
such that
\[
d(\rho(t), T^a \rho_0) + \frac{1}{8\pi} 
\|\nabla V_{\rho(t)}-\nabla V_{T^a \rho_0}\|_2^2 
+ \frac{1}{2} \int |u(t)|^2 \rho(t)\, dx < \epsilon;
\]
here $T^a \rho := \rho(\cdot+a)$.
\end{theorem}

\proof
Assume the assertion were false. 
Then there exist $\e >0$, $t_n>0$, and initial data
$\rho_n(0) \in \F_M$ and $u_n(0)$ such that
for all $n \in \N$, 
\be \label{dist0}
d(\rho_n(0),\rho_0) + \frac{1}{8\pi} \|\nabla V_{\rho_n(0)}-\nabla V_0\|_2^2
+ \frac{1}{2} \int |u_n(0)|^2 \rho_n(0)\, dx < \frac{1}{n}
\ee
but for any $a\in \R^3$,
\be \label{disttn}
d(\rho_n (t_n),T^a \rho_0) + \frac{1}{8\pi} 
\|\nabla V_{\rho_n(t_n)}-\nabla V_{T^a \rho_0}\|_2^2 
+ \frac{1}{2} \int |u_n(t_n)|^2 \rho_n(t_n)\, dx \geq \epsilon.
\ee
By (\ref{dist0}) and (\ref{d-d}),
\[
\lim_{n\to \infty}\H(\rho_n (0),u_n(0)) = \H(\rho_0,0)=\Hr(\rho_0).
\]
Since by assumption the solution $t\mapsto (\rho_n(t),u_n(t))$
conserves energy,
\[
\limsup_{n\to \infty}\Hr( \rho_n (t_n))
\leq \lim_{n\to \infty}\H( \rho_n (t_n),u_n(t_n)) = \Hr(\rho_0),
\]
and since it conserves mass,
$(\rho_n (t_n))\subset \F_M$ is a minimizing sequence for $\Hr$.
By Theorem~\ref{ccp} there exists
a sequence $(a_n) \subset \R^3$ such that up to a subsequence,
\be \label{fieldn}
\|\nabla V_{\rho_n (t_n)}-\nabla V_{T^{a_n} \rho_0}\|_2 \to 0;
\ee
at this point we used the uniqueness of the minimizer.
Note also that for any $\rho \in \F_M$ and $a \in \R^3$,
\[
\|\nabla V_{T^a \rho}-\nabla V_{\rho_0}\|_2
=\|\nabla V_{\rho}-\nabla V_{T^{-a}\rho_0}\|_2,\
d(T^a \rho,\rho_0)=d(\rho,T^{-a}\rho_0).
\]
Since 
$\lim_{n\to \infty}\H(\rho_n (t_n),u_n(t_n)) = \Hr(\rho_0) = \Hr (T^{a_n}\rho_0)$
we conclude by (\ref{fieldn}) and (\ref{d-d}) that
\[
d(\rho_n (t_n),T^{a_n} \rho_0) + 
\frac{1}{2} \int |u_n(t_n)|^2 \rho_n(t_n)\, dx \to 0,\ n \to \infty,
\]
a contradiction to (\ref{disttn}). \prfe

{\bf Remark 4.} Without the uniqueness assumption for the minimizer we
obtain a stability result of the following type:
Let $\M_M \subset \F_M$ denote the set of all minimizers of
$\Hr$ in $\F_M$.
Then for every $\e>0$ there is a $\delta>0$ such that
for any solution
with $\rho(0) \in \F_M$ the initial estimate
\[
\inf_{\rho_0 \in \M_M} 
\left[ 
d(\rho(0),\rho_0) + \frac{1}{8\pi} \|\nabla V_{\rho(0)}-\nabla V_0\|_2^2
+ \frac{1}{2} \int |u(0)|^2 \rho(0)\, dx\right] < \delta
\]
implies that
\be \label{minimsetstab}
\inf_{\rho_0 \in \M_M}
\left[ 
d( \rho (t),\rho_0) + \frac{1}{8\pi} 
\|\nabla V_{ \rho (t)}-\nabla V_{\rho_0}\|_2^2 
+ \frac{1}{2} \int |u(t)|^2 \rho(t)\, dx\right]< \epsilon
\ee
as long as the solution $(\rho(t),u(t))$ exists and preserves mass and energy.
The proof of this assertion follows exactly the same line of reasoning as the 
proof of Theorem~\ref{stability}.

Assume now that the minimizer $\rho_0$ is not necessarily unique,
but isolated up to spatial shifts, that is to say,
\[
\delta_0 := \inf\left\{\| \nabla V_{\rho_0} - \nabla V_{\tilde \rho_0} \|_2
\mid
\tilde \rho_0 \in \M_M\setminus \{T^a \rho_0 \mid a \in \R^3\}\right\}
>0.
\]
In this case the assertion of the theorem again holds,
provided the solution of the Euler-Poisson system is sufficiently
continuous in $t$ so that it cannot jump from one minimizer
to the next, more precisely: 
Let $\epsilon >0$ arbitrary. In order to find the
corresponding $\delta$ we can without loss of
generality assume that
$\epsilon < \delta_0/4$.
Now choose
$\delta > 0$ according to the first part of the present remark so 
that (\ref{minimsetstab}) holds, without
loss of generality $\delta < \epsilon$, and let 
$\rho(0) \in  \F_M$ and $u(0)$ be such that
\[
d(\rho(0),\rho_0) + \frac{1}{8\pi} \|\nabla V_{\rho(0)}-\nabla V_0\|_2^2
+ \frac{1}{2} \int |u(0)|^2 \rho(0)\, dx < \delta .
\] 
The required continuity assumption on the corresponding solution is that
\[
h(t,a)
:=
d(\rho(t), T^a \rho_0) + \frac{1}{8\pi} 
\|\nabla V_{\rho(t)}-\nabla V_{T^a \rho_0}\|_2^2 
+\frac{1}{2} \int |u(t)|^2 \rho(t)\, dx
\]
is continuous, and indeed, that $\inf_{a \in \R^3} h(t,a)$
is continuous. Now
assume that there exists $t>0$ such that
\[
\inf_{a \in \R^3} h(t,a) \geq \epsilon.
\]
Since at time zero the left
hand side is less then $\epsilon$ there exists some
$t^* > 0$ where
\be \label{fndist}
\inf_{a \in \R^3} h(t^\ast,a) = \epsilon.
\ee
On the other hand, the first part of the present remark provides some 
$\rho_0^\ast \in \M_M$ such that
\be \label{infdist}
d( \rho (t^\ast),\rho_0^\ast) + \frac{1}{8\pi} 
\|\nabla V_{ \rho (t^\ast)}-\nabla V_{\rho_0^\ast}\|_2^2 
+ \frac{1}{2} \int |u(t^\ast)|^2 \rho(t^\ast)\, dx < \epsilon \leq
\frac{\delta_0}{4}.
\ee
By (\ref{fndist}) and (\ref{infdist}) together with
the non-negativity of $d$,
\[
\frac{1}{8\pi} 
\|\nabla V_{\rho_0}-\nabla V_{\rho_0^\ast}\|_2^2 \leq \frac{\delta_0}{2},
\]
and by the definition of $\delta_0$ there must exist some
$a^\ast \in \R^3$ such that
$\rho_0^\ast = T^{a^\ast} \rho_0$.
But this means that (\ref{fndist}) contradicts (\ref{infdist}),
and the argument is complete.
\smallskip

{\bf Remark 5.}
The restriction $\rho(0) \in \F_M$ for the perturbed initial data is
acceptable from a physics point of view: A small perturbation
of a given star, say by the gravitational pull of some outside object,
results in a perturbed state with the same mass.
However, such a perturbation will hardly be spherically symmetric
so that it is important that no such restriction is necessary in our 
stability result. On the other hand, if we restrict ourselves to 
spherically symmetric perturbations then the spatial shifts in the
statement of the stability result are no longer necessary.
Their need arises from the fact
that a given steady state can be given a uniform velocity
in one direction so that it drifts off and the distance of the perturbed
state from the original one grows linearly in $t$, no matter how
small the initial perturbation was.

\section{Fluid models versus kinetic models}
\setcounter{equation}{0}

Obviously, the hard part in the analysis, if any,
is the proof of Theorem~\ref{ccp}.
The essential arguments for that theorem as well
as its exploitation for stability questions arose from
the investigation of the stability of galaxies by Y.~Guo
and the author. In typical galaxies 
even stars which are spatially close to
each other can have very different velocities, and hence galaxies
are modeled not by fluid equations but
by kinetic equations, i.e., by the Vlasov-Poisson system
which was stated in the introduction.
On the level of existence and regularity of solutions
to the initial value problem the Vlasov-Poisson
system is very different from the Euler-Poisson system:
Continuously differentiable and
compactly supported initial data for $f$ launch continuously
differentiable solutions which are global in time, cf.\ \cite{Pf,LP,Sch}, 
and neither shocks nor a gravitational collapse can occur.
On the other hand, as to the existence and stability of stationary
solutions both systems seem to be intimately related,
and in this last section we want to comment on this relation.

In \cite{G1,G2,GR1,GR2,R1,R2} steady states of the Vlasov-Poisson system
were obtained as minimizers
of an energy-Casimir functional
\[
\Hc(f)=\int\!\!\!\!\int Q(f(x,v))\,dv\,dx 
+ \frac{1}{2} \int\!\!\!\!\int |v|^2 f(x,v)\,dv\,dx
- \frac{1}{2} \int\!\!\!\!\int \frac{\r_f(x)\, \r_f(y)}{\n{x-y}}
\, dx\, dy
\]
on the constraint set
\[
\G_M := \Bigl\{ f \in L^1 (\R^6) 
\mid
f\geq 0,\ \C(f) + \ekin(f) < \infty,\ 
\ \int\!\!\!\!\int f\,dv\,dx = M \Bigr\},
\]
where $\C(f)$ and $\ekin(f)$ denote the first and second 
term in the energy-Casimir functional respectively. Such minimizers exist,
satisfy assertions analogous to Theorems~\ref{ccp} and \ref{minim=ss},
and are non-linearly stable, provided $Q$ satisfies assumptions
which are exactly parallel to the ones for $\Phi$, except that
the parameters $0<n,n'<3$ have to be replaced by parameters
$0<k,k'<3/2$.

In \cite{R3} it was observed that there is a one-to-one
correspondence between the minimizers of the energy-Casimir
functional $\Hc$ on the ``kinetic'' set $\G_M$ and the minimizers
of the ``reduced'' functional $\Hr$ on the set $\F_M$:
If $\rho_0$ is a minimizer of the latter functional with
induced potential $V_0$ then 
\[
f_0
:=
\left\{ 
\begin{array}{ccl}
(Q')^{-1}(E_0 - E) &,& E < E_0, \\
0 &,& E \geq E_0 
\end{array}
\right.
\]
is a minimizer of $\Hc$, where the particle energy $E$ is defined
as
\[
E = E(x,v):=\frac{1}{2} |v|^2 + V_0(x).
\] 
On the other hand, if $f_0$ is a minimizer of $\Hc$ then the induced
spatial density $\rho_0=\int f_0 dv$ minimizes $\Hr$.
Of course, in order for this correspondence to be true the
functions $Q$ and $\Phi$ have to be in the proper relation:
\[ 
\Phi(r)=\inf_{g \in\G_r} 
\int\left(\frac{1}{2}|v|^2 g (v) + Q(g(v))\right)\,dv
\]
where for $r\geq 0$,
\[
\G_r :=\left\{g \in L^1 (\R^3) | g\geq 0,\ 
\int\left(\frac{1}{2}|v|^2 g (v) + Q(g(v))\right)\,dv < \infty,\
\int g(v)\, dv = r \right\}. 
\]
With this relation it follows that for every $f \in \G_M$,
\[
\Hc(f) \geq \Hr(\r_f),
\]
and if $f=f_0$ is a minimizer of $\Hc$ over $\F_M$ then equality holds;
one first minimizes over all $f$ which give the same spatial density $\rho$
and then minimizes over the latter.
Indeed, there is a more explicit relation between $Q$
and $\Phi$ via their Legendre transforms;
for a function 
$h:\R \to ]-\infty,\infty]$ we denote by
\[
h^\ast (\lambda):= \sup_{r \in \R} (\lambda \, r - h(r))
\]
its Legendre transform. If we extend $\Phi$ and $Q$ to $]-\infty,0[$
by $+\infty$ then
\[
\Phi^\ast (\lambda)
=
\int Q^\ast \left( \lambda - \frac{1}{2} |v|^2 \right)\, dv,
\]
for $\lambda \in \R$,  
and it can be seen that the assumptions
on $Q$ translate into the corresponding ones for $\Phi$
where the relevant exponents are related by $n=k+3/2$.
These assertions are established in \cite{R3}, cf.\ also \cite{W}.

If one is interested only in the existence of 
spherically symmetric steady states
and not their stability the relation becomes even more direct:
Suppose $f_0$ is such a steady state of the Vlasov-Poisson system
and define its radial and tangential pressures as functions of the radial
variable $r=|x|$ by
\[
p_0(r)=\int\left(\frac{x\cdot v}{r}\right)^2 f_0(x,v)\, dv,\ 
p_0^T(r)=\frac{1}{2}\int\left|\frac{x\times v}{r}\right|^2 f_0(x,v)\, dv. 
\]
Then a simple computation using the Vlasov equation implies the relation
\be \label{tov}
p_0' = \frac{2}{r}(p_0^T - p_0) - \rho_0 V_0',
\ee
the Tolman-Oppenheimer-Volkov equation. If we make the isotropic ansatz 
\be \label{isotropic}
f_0(x,v) = \phi (E(x,v))
\ee
with $\phi\geq 0$ prescribed and the particle energy $E$ given in terms
of the potential $V_0$ as above then the pressure is isotropic,
$p_0^T = p_0$, and (\ref{tov}) reduces to the static
Euler equation $p_0'+\rho_0 V_0'=0$. Moreover,
\beas
\rho_0(r) 
&=& 2^{5/2} \pi
\int_{V_0(r)}^\infty \phi(E)\,(E-V_0(r))^{1/2}\, dE =: g_\phi(V_0(r)),\\
p_0 (r)
&=&
\frac{2^{7/2}}{3} \pi
\int_{V_0(r)}^\infty \phi(E)\,(E-V_0(r))^{3/2}\, dE =: h_\phi(V_0(r)).
\eeas
Hence the ansatz (\ref{isotropic}) reduces the stationary Vlasov-Poisson 
system to the semilinear Poisson equation
\be \label{semilinp}
\lap V_0 = 4 \pi g_\phi(V_0).
\ee
A solution of the latter
automatically induces a solution of the
static Euler-Poisson system which equation of state
$p_0=h_\phi\circ g_\phi^{-1}(\rho_0)$; both $g_\phi$ and $h_\phi$
are invertible on their support under
mild assumptions on $\phi$, cf.\ \cite{RR}.
On the other hand, starting from the static, spherically symmetric 
Euler-Posson system one can integrate the Euler equation using the equation
of state and express $\rho_0$ as a function of the potential
$V_0$ so that again the problem is reduced to a semilinear
Poisson equation, which is precisely (\ref{semilinp})
if the equation of state has the form $p_0=h_\phi\circ g_\phi^{-1}(\rho_0)$. 

However, these parallels between static solutions of the two systems
break down at several points:
First of all, static solutions of the Euler-Poisson system,
that is, steady states with vanishing velocity field, must be spherically
symmetric, cf.\ \cite{Lich}, but for the Vlasov-Poisson
system there exist axially symmetric steady states
which are not spherically symmetric
and which have non-vanishing velocity field $\int v f_0 dv \neq 0$,
cf.\ \cite{R4}. 

Another point where the two systems differ, now concerning 
the question of stability, is the following: For the Euler-Poisson
system the threshold $n<3$ which corresponds to $\gamma >4/3$
for the equation of state is sharp, since for the polytropic case
with $\gamma=4/3$
the energy of a steady state is zero and an arbitrarily small
perturbation of such a state can make the energy positive and
cause part of the system to travel off to infinity.
These assertions are shown in \cite[Thm.~1.3 (iv), Thm.~1.4]{DLYY}.
For the Vlasov-Poisson system one can go beyond this threshold to
obtain stability
provided $0<k<7/2$ which corresponds to $3/2 < n=k+3/2 < 5$.
This was done in \cite{GR3} by minimizing the energy $\ekin + \epot$
under the mass-Casimir constraint $\int f + \C(f) = M$.
Moving the Casimir  functional $\C$ from the functional which we minimize into
the constraint allows for an extension of the stability result
which includes all the polytropes with index up to and including 
$n=k+3/2 = 5$. The reduction mechanism which takes us
to the Euler-Poisson system does no longer
work in this context, $\int \Phi(\rho)\,dx$ is not conserved
by itself so that we cannot move this functional from the minimized 
functional into the constraint, and all this fits with the fact that---as 
opposed to the Vlasov-Poisson system---the threshold $n<3$ is sharp 
for the Euler-Poisson system as regards stability.

To conclude we note that if the variational formulation is chosen 
as in \cite{GR3} then one can prove the stability
result in the form stated in Theorem~\ref{stability}
without assuming uniqueness or even isolatedness of the minimizer,
cf.\ \cite{Sch2}. Whether this is possible also in the
framework of the present note is one of the many open problems
in this area.  
\bigskip

\noindent
{\bf Acknowledgment.}
This paper originates from my collaboration with Y.~Guo, Brown University,
whom I sincerely thank for many stimulating discussions.
The research was supported by the Wittgenstein 
2000 Award of P.~A.~Markowich.


\begin{thebibliography}{10}

\bibitem{BT}
{\sc Binney, J., Tremaine, S.}:
{\em Galactic Dynamics},
Princeton University Press, 1987

\bibitem{Chan}
{\sc Chandrasekhar,S.}:
{\em Hydrodynamic and Hydromagnetic stability},
Dover, 1981

\bibitem{DLYY}
{\sc Deng, Y., Liu, T.-P., Yang, T. \& Yao, Z.-A.}:
Solutions of Euler-Poisson equations for gaseous stars.
{\em Arch.\ Rational Mech.\ Anal.}, to appear

\bibitem{Duc}
{\sc Ducomet, B.}:
Hydrodynamical models of gaseous stars.
{\em Review of Mathematical Physics}
{\bf 8}, 957--1000 (1996)

\bibitem{GNN}
{\sc Gidas, B., Ni, W.-M., Nirenberg, L.}:
Symmetry and related properties via the maximum principle.
{\em Commun.\ Math.\ Phys.}\
{\bf 68}, 209--243 (1979)


\bibitem{G1}
{\sc Guo, Y.}:
Variational method in polytropic galaxies.
{\em Arch.\ Rational Mech.\ Anal.},
{\bf 150}, 209--224 (1999)

\bibitem{G2}
{\sc Guo, Y.}:
On the generalized Antonov's stability criterion.
{\em Contem. Math.} {\bf 263}, 85--107 (2000)

\bibitem{GR1}
{\sc Guo, Y., Rein, G.}:
Stable steady states in stellar dynamics.
{\em Arch.\ Rational Mech.\ Anal.}\
{\bf 147}, 225--243 (1999)

\bibitem{GR2}
{\sc Guo, Y., Rein, G.}:
Existence and stability of Camm type steady states in galactic 
dynamics.
{\em Indiana University Math.\ J.},
{\bf 48}, 1237--1255 (1999)

\bibitem{GR3}
{\sc Guo, Y., Rein, G.}:
Isotropic steady states in galactic dynamics.
{\em Commun.\ Math.\ Phys. }, 
{\bf 219}, 607--629 (2001)

\bibitem{GR4}
{\sc Guo, Y., Rein, G.}:
Stable models of elliptical galaxies.
Preprint 2002,
 
\bibitem{HMRW}
{\sc Holm, D.~D., Marsden, J.~E., Ratiu, T., \& Weinstein, A.}:
Nonlinear stability of fluid and plasma equilibria.
{\em Physics Reports}, 
{\bf 123}, Nos.\ {\bf 1} and {\bf 2}, 1--116  (1985)

\bibitem{Kipp}
{\sc Kippenhahn, R., Weingert, A.}:
{\em Stellar Structure and Evolution},
Springer Verlag, 1994

\bibitem{Lich}
{\sc Lichtenstein, L.}:
{\em Gleichgewichtsfiguren rotierender Fl\"ussigkeiten},
Springer, Berlin, 1933

\bibitem{LL}
{\sc Lieb, E.~H., Loss, M.}:
{\em Analysis}.
American Mathematical Society, Providence 1996

\bibitem{Lin}
{\sc Lin, S.-S.}:
Stability of gaseous stars in spherically symmetric motions.
{\em SIAM J.\ Math.\ Anal.},
{\bf 28}, 539--569 (1997)

\bibitem{L}
{\sc Lions, P.-L.}: 
The concentration-compactness principle in the calculus 
of variations.
The locally compact case. Part 1.
{\em Ann.\ Inst.\ H.\ Poincar\'{e}}
{\bf 1}, 109--145 (1984)

\bibitem{LP}
{\sc Lions, P.-L., Perthame, B.}:
Propagation of moments and regularity for the 3-dimensional
Vlasov-Poisson system.
{\em Invent.\ Math.}\ {\bf 105}, 415--430 (1991).

\bibitem{Pf}
{\sc Pfaffelmoser, K.}:
Global classical solutions of the Vlasov-Poisson system in three
dimensions for general initial data.
{\em J.\ Diff.\ Eqns.}\ 
{\bf 95}, 281--303 (1992)

\bibitem{R1}
{\sc Rein, G.}:
Flat steady states in stellar dynamics---existence and stability.
{\em Commun.\ Math.\ Phys.}\
{\bf 205}, 229--247 (1999)

\bibitem{R4}
{\sc Rein, G.}:
Stationary and static stellar dynamic models with axial symmetry. 
{\em Nonlinear Analysis; Theory, Methods \& Applications}
{\bf 41}, 313--344 (2000)

\bibitem{R2}
{\sc Rein, G.}:
Stability of spherically symmetric steady states in galactic 
dynamics against general perturbations.
{\em Arch.\ Rational Mech.\ Anal.},
{\bf 161}, 27--42 (2002)

\bibitem{R3}
{\sc Rein, G.}:
Reduction and a concentration-compactness principle
for energy-Casimir functionals,
{\em SIAM J.\ on Mathematical Analysis}, 
{\bf 33}, 896--912 (2002)

\bibitem{RR}
{\sc Rein, G., Rendall, A.~D.}:
Compact support of spherically symmetric equilibria in 
non-relativistic and relativistic galactic dynamics.
{\em Math.\ Proc.\ Camb.\ Phil.\ Soc.}\ 
{\bf 128}, 363--380 (2000)

\bibitem{Sch}
{\sc Schaeffer, J.}:
Global existence of smooth solutions to the Vlasov-Poisson system
in three dimensions.
{\em Commun.\ Part.\ Diff.\ Eqns.}\ 
{\bf 16}, 1313--1335 (1991)

\bibitem{Sch2}
{\sc Schaeffer, J.}:
On steady states in galactic dynamics.
Preprint, 2002


\bibitem{W}
{\sc Wolansky, G.}:
On nonlinear stability of polytropic galaxies.
{\em Ann.\ Inst.\ Henri Poincar\'{e}},
{\bf 16}, 15--48 (1999)

\end{thebibliography}
\end{document}